\documentclass{article}
\usepackage{cite,bm}
\usepackage{spconf,amsmath,graphicx}
\usepackage{xcolor}
\usepackage{booktabs}
\usepackage{multirow}
\usepackage{hyperref}
\usepackage{setspace}
\usepackage{amssymb}

\newcommand{\etal}{\textit{et al.}}

\title{Exploring Speech Enhancement for Low-resource Speech Synthesis}

\name{
\begin{tabular}{@{}c@{}}
Zhaoheng Ni$^{1}$\qquad Sravya Popuri$^{1}$\qquad Ning Dong$^{1}$\qquad Kohei Saijo$^{2}$\\ Xiaohui Zhang$^{1}$\qquad Gael Le Lan$^{1}$\qquad Yangyang Shi$^{1}$\qquad Vikas Chandra$^{1}$\qquad Changhan Wang$^{1}$
\end{tabular}}

\address{$^{1}$Meta AI, USA\qquad $^{2}$Waseda University, Japan}

\begin{document}
\maketitle
\begin{abstract}
High-quality and intelligible speech is essential to text-to-speech (TTS) model training, however, obtaining high-quality data for low-resource languages is challenging and expensive. Applying speech enhancement on Automatic Speech Recognition (ASR) corpus mitigates the issue by augmenting the training data,  while how the nonlinear speech distortion brought by speech enhancement models affects TTS training still needs to be investigated. In this paper, we train a TF-GridNet speech enhancement model and apply it to low-resource datasets that were collected for the ASR task, then train a discrete unit based TTS model on the enhanced speech. We use Arabic datasets as an example and show that the proposed pipeline significantly improves the low-resource TTS system compared with other baseline methods in terms of ASR WER metric. We also run empirical analysis on the correlation between speech enhancement and TTS performances.
\end{abstract}
\section{Introduction}
\label{sec:intro}

Speech synthesis or text-to-speech (TTS) aims at generating high-quality and intelligible human speech given the text transcription. Deep learning based TTS systems have dominated over conventional approaches such as concatenative synthesis~\cite{hunt1996unit} and statistical parametric synthesis~\cite{black2007statistical}. Deep learning based TTS systems can be categorized into two categories based on the generation style: 1) Autoregressive methods~\cite{li2019neural, wang2017tacotron, shen2018natural} which generate mel-spectrogram autoregressively given input text then generates waveform via vocoders such as  Griffin-Lim~\cite{griffin1984signal}, WaveGlow~\cite{prenger2019waveglow}, WaveRNN~\cite{kalchbrenner2018efficient}, etc. 2) Non-autoregressive methods which generates mel-spectrogram in parallel~\cite{kim2020glow, ren2019fastspeech, ren2020fastspeech}. Recently, Eloff \etal~replaced the mel-spectrogram input with a speech representation learnt by a VQ-VAE model and pass it to an FFTNet vocoder to generate speech waveforms~\cite{eloff2019unsupervised}. Polyak \etal~\cite{polyak2021speech} improved the speech representations by extracting discrete units from self-supervised learning (SSL) models and pass them to a HiFiGAN vocoder~\cite{kong2020hifi}.

Although there is significant progress on TTS research, there are still many challenges for low-resource languages. For example, unlike high-resource languages such as English, German, and Mandarin, it is difficult to obtain high-quality recordings for low-resource languages such as Arabic, Mongolian, Cebuano. On the other hand, there are huge amount of data for low-resource languages that have been collected for other purposes (e.g. Automatic Speech Recognition (ASR)). Studies~\cite{ogun2023can, cooper2017utterance} trained TTS models on ASR corpus and show that selecting intelligible speech for training improved the TTS performance. However, due to the noisy characteristics of the training data, the generated speech quality still has a gap compared to TTS models trained on high-quality datasets.

Speech enhancement approaches can improve the speech quality and intelligibility by reducing background noises and interfering speakers while preserving the target speech signal. Lu~\etal~showed that applying speech enhancement benefit other downstream tasks such as speech translation~\cite{lu2022espnet}. Wang \etal~applied a speech enhancement model on mel-spectrogram based TTS models and significantly improved the TTS performance in terms of mean opinion score (MOS) and ASR Word Error Rate (WER)~\cite{wang2021fairseq}. However, this work did not study the choice of speech enhancement models.

Although speech enhancement approaches are promising to mitigate the issue of TTS training on ASR corpus, \cite{du2014robust, xu2020neural} showed that single-channel speech enhancement models may introduce nonlinear speech distortions that downgrades ASR performances. For low-resource TTS models, how different single-channel speech enhancement models affect the model training has not yet been investigated.

\begin{figure*}[t]
    \centering
    \includegraphics[width=\textwidth]{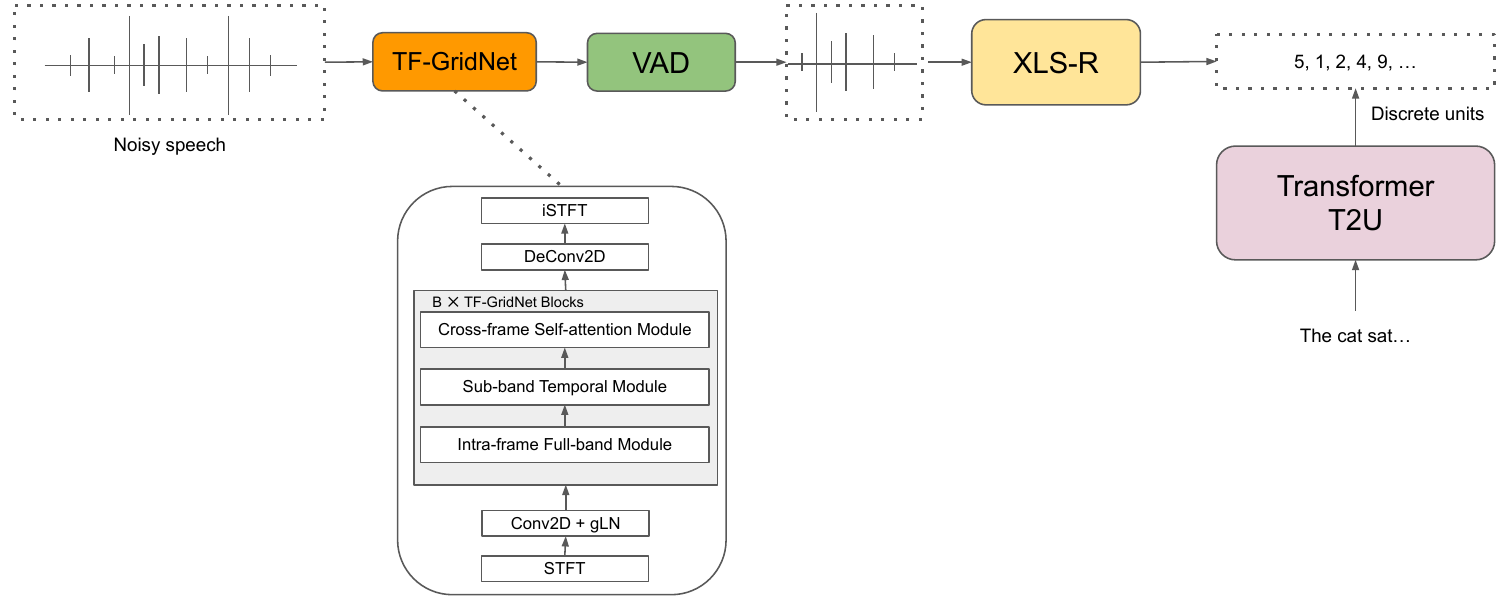}
    \caption{The diagram of our proposed system. We first apply the TF-GirdNet model to reduce the noises, then trim the enhanced speech with a VAD model. We use the XLS-R model to extract the transformer representations then extract the discrete units by k-means clustering, with which we train a text-to-unit model with the text input and corresponding discrete units.}
    \label{fig:pipeline}
\end{figure*}

    To answer this question, we trained a TF-GridNet~\cite{wang2023tf} based speech enhancement model and apply it onto Arabic ASR datasets and trained a text-to-unit (T2U) model as in SeamlessM4T~\cite{communication2023seamlessm4tmassively}. The generated discrete units are passed to a unit based HiFiGAN vocoder to generate waveforms. Compared with other state-of-the-art speech enhancement systems (DEMUCS~\cite{defossez2020real}, FullSubNet~\cite{hao2021fullsubnet}, FRCRN~\cite{zhao2022frcrn}, and D2Former~\cite{zhao2023d2former}), the results show TF-GridNet significantly improves T2U performance in terms of the ASR WER metric.
    We conduct empirical analysis on correlation between speech enhancement and TTS performances and find that high speech enhancement performance doesn't always align with T2U model performance.

The remained of the paper is structured as follows. Section~\ref{sec:method} introduces the proposed pipeline for applying speech enhancement for low-resource TTS training. We present the datasets and experiments setup in Section~\ref{sec:experiments}. Results are reported and analyzed in Section~\ref{sec:results}. We draw conclusions in Section~\ref{sec:conclusion}

\section{Methods}
\label{sec:method}
\subsection{Problem Formulation}
A noisy speech signal ${\bf{x}} \in \mathbb{R}^{L}$
can be modeled as:
\begin{equation}
    \bf{x}_i = \bf{s}_i + \bf{n}_i,
\end{equation}
where $\bf{s}_i$ refers to the clean speech and $\bf{n}_i$ refers to the addictive noise for frame $i$.
Given such speech signal $\bf{x}$ and the corresponding text transcription $\bf{t}$, our goal is to build a TTS system $\mathcal{G}$ that takes $\bf{t}$ as input and generates speech that is as clean and intelligible as the clean speech $\bf{s}$.

Figure~\ref{fig:pipeline} shows the pipeline of our designed system. The noisy speech $\bf{x}$ is processed by a speech enhancement model named as TF-GridNet which is described in Section~\ref{subsec:tfgridnet} followed by a voice activity detection (VAD) model to get $\hat{\bf{s}}$. Then we extract self-supervised learning (SSL) features via the XLS-R-1B model~\cite{babu2021xls} which is a SSL model with one billion parameters pre-trained on 128 languages. We use the 35$^{th}$ transformer layer's output as the SSL feature and convert it to discrete units by k-means clustering algorithm. Then we train a text-to-unit model which is described in Section~\ref{subsec:t2u} that predicts the discrete unit sequence given the input text. The predicted discrete units are converted to waveform by a discrete unit based HiFiGAN vocoder.

\subsection{Text-to-Unit}
\label{subsec:t2u}
The text-to-unit (T2U) model is a Transformer-based encoder-decoder model trained on aligned text and discrete units from ASR data. We use the Transformer model in fairseq~\footnote{\url{https://github.com/facebookresearch/fairseq}}
for T2U model training. The model has an encoder-encoder architecture with 6 encoder layers and 6 decoder layers respectively. We refer the readers to the Section 4.3 in SeamlessM4T~\cite{communication2023seamlessm4tmassively} for more details on T2U model and unit extraction.
As autoregressive models do not work well on predicting repeating tokens, we reduce the repeating units by checking if they appear in consecutive frames.

\subsection{TF-GridNet}
\label{subsec:tfgridnet}

Speech enhancement aims to extract clean speech signal $\bf{s}$ from noisy speech $\bf{x}$ by removing noise $\bf{n}$. We utilize the speech enhancement technique to improve the T2U performance by removing noise $\bf{n}$ from the noisy dataset. To this end, we exploit TF-GridNet~\cite{wang2023tf}, a recently proposed speech enhancement model that achieved high performance on many benchmarks. The TF-GridNet utilizes the short-time Fourier transform (STFT) and inverse STFT (iSTFT) as the encoder and decoder and estimates the real and imaginary parts of the target speech via complex spectral mapping. In the following, we briefly describe the model architecture, which is shown in Figure~\ref{fig:pipeline}. 

In TF-GridNet, the input noisy speech $\bf{x}$ is first transformed into a time-frequency domain feature with a shape of ${2 \times T \times F}$ via STFT, where 2 denotes the real and imaginary parts. The STFT feature is then mapped into the D-dimensional space by 2D convolution with a $3 \times 3$ kernel followed by a global layer normalization (gLN)~\cite{luo2019conv}. The resulting feature with a shape of ${D \times T \times F}$ is fed into $B$ TF-GridNet blocks, each of which is composed of an intra-frame full-band module, a sub-band temporal module, and a cross-frame self-attention module. The intra-frame full-band module consists of an LN, an unfolding process, a bi-directional long short-term memory (BLSTM), and a 1D deconvolution. The input is seen as $T$ separate $D$-dimensional embeddings with sequence length $F$, and the core processing part, BLSTM, models frequency information in each frame. The sub-band temporal module has the same architecture but it sees the input as $F$ separate $D$-dimensional embeddings with sequence length $T$ to model temporal information. In contrast to these modules which model frequency or temporal information step by step using BLSTM, the final cross-frame self-attention module is designed to capture long-context information to enable each frame information to be directly reflected in any other frames. After processing with $B$ TF-GridNet blocks, a 2D deconvolution with a $3 \times 3$ kernel is applied to obtain the estimate of the real and imaginary parts with a shape of ${2 \times T \times F}$. Finally, the time-domain enhanced speech signal is obtained via iSTFT.

Although the original TF-GridNet estimates the target speech directly by complex spectral mapping, we found that, in our case, complex spectral mapping sometimes generates unexpected noises at the end of each utterance. We mitigate this issue by training the model to estimate the complex ratio mask~\cite{williamson2015complex} instead.

\subsection{Voice Activity Detection}
VAD models aims to detect whether there is an active speaker in the frames, which can be regarded as a binary classification task. After speech enhancement, the noise regions will become silence, which is redundant for TTS training. By applying VAD models to the enhanced speech and trimming the silent regions, the signal is shorter which helps TTS model faster to converge.

\section{Experiments}
\label{sec:experiments}
\subsection{Datasets}
We use the DNS 2020 challenge dataset~\cite{reddy2020interspeech}  for training the TF-GridNet model. The clean speech dataset contains over 500 hours of utterances from 2150 speakers. The noise dataset contains over 180 hours of clips from 150 classes. Different from the official setting, we simulate the noisy speech with a lower random signal-to-noise ratio (SNR) in range [-35, -15] to make the model more robust to noises. We use four Arabic datasets for T2U training: Arabic subset of Common-Voice~\cite{commonvoice:2020}, Massive Arabic Speech Corpus (MASC)~\cite{al2023masc}, the Arabic subset of Multilingual TEDx corpus~\cite{salesky2021multilingual}, and the Arabic subset of MediaSpeech corpus~\cite{kolobov2021mediaspeech}. The total training data consists of 890 hours of Arabic speech and 764,929 utterances. For evaluation we use the Arabic test-set of FLEURS~\cite{fleurs2022arxiv} dataset which contains 428 utterances. 

\begin{table}[t]
    \centering
    \small
    \begin{tabular}{c|c}
        \toprule
        Model & WER (\%) $\downarrow$ \\
        \midrule
        Unprocessed &  57.6 \\
        DEMUCS + VAD & 46.6 \\
        \bottomrule
    \end{tabular}
    \caption{Ablation study on the speech enhancement and VAD components to the T2U model training.}
    \label{tab:ablation}
\end{table}

\begin{table*}[t]
    \centering
    \small
    \begin{tabular}{l|ccc|ccc}
    \toprule
        \multirow{2}{*}{Model} & \multicolumn{3}{c|}{Without Reverb} 
 & \multicolumn{3}{c}{With Reverb} \\
 \cline{2-7}
 & WB-PESQ $\uparrow$& Si-SDR (dB) $\uparrow$ & STOI (\%) $\uparrow$ & WB-PESQ $\uparrow$ & Si-SDR (dB) $\uparrow$ & STOI (\%) $\uparrow$ \\
    \midrule
        DEMUCS~\cite{defossez2020real} &  2.7 & 19.0 & 96.5 & 1.6 & 6.0& 77.8\\
        FullSubNet~\cite{hao2021fullsubnet} & 2.9 & 17.6 & 96.4 & \textbf{3.0} & \textbf{15.8}& \textbf{92.7}\\
        FRCRN~\cite{zhao2022frcrn} & \textbf{3.2} & 20.0 & \textbf{97.7} & 1.8& 9.4& 85.5\\
        D2Former~\cite{zhao2023d2former} & 2.5 & 15.0 & 94.8 &1.2 & 1.6& 60.5\\
        TF-GridNet & \textbf{3.2} & \textbf{20.6} & \textbf{97.7} & 2.8 & 14.0& 90.3\\
    \bottomrule
    \end{tabular}
    \caption{Comparisons with other speech enhancement models in terms of wideband PESQ, STOI, and Si-SDR on the DNS 2020 challege test sets.}
    \label{tab:se-metrics}
\end{table*}

\subsection{Training}
We train the TF-GridNet model by modifying the recipe~\footnote{\url{https://github.com/espnet/espnet/tree/master/egs2/dns_ins20/enh1}} in ESPNet toolkit~\cite{lu2022espnet}. The model has FFT size of 512, hop length of 256, $D$ of 32, $B$ of 4, LSTM hidden size of 128. We use the scale-invariant signal-to-distortion ratio (Si-SDR)~\cite{le2019sdr} as the loss function and train the model for 30 epochs.
We use 2 Nvidia Tesla P100 GPUs with a total batch size of 15. For DEMUCS~\footnote{\url{https://github.com/facebookresearch/DEMUCS}}, FullSubNet~\footnote{\url{https://github.com/Audio-WestlakeU/FullSubNet}}, and D2Former~\footnote{\url{https://github.com/alibabasglab/D2Former}} we use the shared pre-trained checkpoints from GitHub. For FRCRN we use the pre-trained pipeline in the ModelScope library~\footnote{\url{https://github.com/modelscope/modelscope}}. We apply the same webrtcvad~\footnote{\url{https://github.com/wiseman/py-webrtcvad}} model for voice activity detection after different speech enhancement models.

For text-to-unit model training, we train the Transformer to 200K updates. Each layer has 16 attention heads, embedding size of 1024, FFN size of 8192, attention dropout rate of 0.1. The initial learning rate is linearly warmed up to 1e-3 in the first 4000 updates, then decayed with the rate of inverse square root of the iteration number. We use the cross entropy loss function with a label smoothing parameter of 0.1. We train each model on 8 Nvidia Tesla V100 GPUs with maximum 2048 tokens in a batch.

\subsection{Evaluation}
For evaluationg the speech enhancement models, we use the evaluation script~\footnote{\url{https://github.com/espnet/espnet/blob/master/espnet2/bin/enh_scoring.py}} in ESPNet toolkit. We estimate the enhanced speech and compute the wideband PESQ (WB-PESQ)\cite{rix2001perceptual}, Si-SDR metrics which measure the speech quality
and Short Term Objective Intelligibility (STOI)~\cite{taal2011algorithm} metric which measures the speech intelligibility.

For evaluating the T2U model, we synthesize the speech by passing the text input to the T2U model and pass the estimated discrete units to a HiFiGAN vocoder which is trained on units of 36 languages~\cite{communication2023seamlessm4tmassively}. In terms of evaluation metric, we use the Whisper Large-V2 ASR model~\cite{radford2023robust} to transcribe the synthetic speech and calculate WER with the reference transcription. We choose the checkpoint that has the lowest WER on the dev set of FLEURS dataset and report the WER on the test set.

\section{Results}
\label{sec:results}
To verify the effectiveness of the proposed pipeline, we first apply the DEMUCS model and VAD to the noisy speech, then train a T2U model with the enhanced units. Table~\ref{tab:ablation} shows compared to the model trained on unprocessed speech, the proposed pipeline shows significant improvement in terms of WER metric.

Table~\ref{tab:se-metrics} shows the speech enhancement performance of TF-GridNet on DNS 2020 test sets. Note that D2Former is trained on VoiceBank+Demand dataset, while the others are trained on the DNS 2020 dataset. Compared with other baselines, TF-GridNet achieves the best performance on the non-revebrant test set in terms of wideband PESQ, Si-SDR, and STOI. FullSubNet achieves the best performance on reverbrant test set since they mix the training data with random room impulse responses~\cite{hao2021fullsubnet}. Although TF-GridNet doesn't see any reverbrant speech data, it shows significant improvements on the reverbrant test set compared to other methods except FullSubNet. D2Former doesn't perform well on the DNS 2020 test sets, which indicates the training and test data mismatch affects the performance significantly.

Table~\ref{tab:se-wer} shows the WER results of the original FLEURS test-set (Reference), the speech which is synthesized by the discrete units of the original speech (Re-synthesis), the speech synthesized by the T2U model trained on unprocessed data (Unprocessed), and thoese by the T2U models trained on the enhanced speech. The results show that the T2U model trained on TF-GridNet enhanced speech achieves the lowest WER score with a 12.9\% absolute reduction compared to the T2U model trained on unprocessed speech. The audio of FLUERS dataset is also noisy, hence the WER of re-synthesized speech is close to "Unprocessed". All T2U models trained on enhanced speech surpass the "Re-synthesis" and "Unprocessed", indicating the importance of speech enhancement in the pipeline. Although FullSubNet performs the best in terms of speech enhancment on the reverbrant speech, it doesn't outperform other systems except D2Former. We analyze the Arabic training data and find there is few reverberation in the speech, which indicates the data mismatch between the speech enhancement model and the T2U training data causes the performance gap. Same for D2Former which is trained on a smaller dataset, the generalization ability is limited.

\begin{table}[t]
    \centering
    \small
    \begin{tabular}{l|c}
    \toprule
        Model& WER (\%) $\downarrow$ \\
    \midrule
        Reference & 18.6 \\
        Re-synthesis & 57.4 \\
        \toprule
        Unprocessed & 57.6 \\
        DEMUCS~\cite{defossez2020real} &  46.6 \\
        FullSubNet~\cite{hao2021fullsubnet} &  47.1\\
        FRCRN~\cite{zhao2022frcrn} & 46.2\\
        D2Former~\cite{zhao2023d2former} & 50.6\\
        TF-GridNet & \textbf{44.7} \\
    \bottomrule
    \end{tabular}
    \caption{T2U performance in terms of ASR WER metric on FLEURS test set.}
    \label{tab:se-wer}
\end{table}

\section{Conclusion}
\label{sec:conclusion}
We train a TF-GridNet speech enhancement model and design a text-to-unit pipeline for low-resource speech synthesis. The proposed approach significantly improves the T2U performance compared to other speech enhancement models. In our future work, we will focus on designing a better metric of speech enhancement models that can reflect the downstream low-resource speech synthesis performance.

\bibliographystyle{IEEEbib}
\small
\setstretch{0.9}
\bibliography{refs}

\end{document}